\documentclass[twoside,amsmath,amssymb,superscriptaddress,showpacs,nofootinbib,aps]{revtex4-1}
\usepackage{graphicx}
\usepackage{float}

\usepackage{slashed}
\usepackage{amsmath}
\usepackage{amsthm}
\usepackage{subfigure}
\usepackage[utf8]{inputenc}
\usepackage{graphicx}
\usepackage{epsfig}
\usepackage{amssymb}
\usepackage{dcolumn}
\usepackage{bm}

\newcommand{\ba}{\begin{array}}
\newcommand{\ea}{\end{array}}
\def\br{\begin{eqnarray}}
\def\er{\end{eqnarray}}
\def\be{\begin{equation}}
\def\ee{\end{equation}}
\usepackage{hyperref}

\def\({\left(}
\def\){\right)}

\def\<{\left\langle}
\def\>{\right\rangle}

\newcommand{\Q}{\textnormal{\tiny \textsc{Q}}}
\newcommand{\T}{\textnormal{\tiny \textsc{T}}}
\newcommand{\E}{\textnormal{\tiny \textsc{E}}}
\def\tt{\textnormal\tiny\textsc}

\begin{document}

%
%
\title{Technicolor coupled models }
%
%
\author{A. A. Natale}
\email{adriano.natale@unesp.br}
\affiliation{Instituto de F{\'i}sica Te\'orica - UNESP, Rua Dr. Bento T. Ferraz, 271,\\ Bloco II, 01140-070, S\~ao Paulo, SP, Brazil}

\author{A. Doff}
\email{agomes@utfpr.edu.br}
\affiliation{Universidade Tecnol\'ogica Federal do Paran\'a - UTFPR - DAFIS \\
Av Monteiro Lobato Km 04, 84016-210, Ponta Grossa, PR, Brazil}

%


\begin{abstract}
When technicolor (TC), QCD, extended technicolor (ETC) and other interactions become coupled through their different Schwinger-Dyson equations, the solution of these equations are modified compared to those of the isolated equations. The change in the self-energies is similar to that obtained in the presence of four-fermion interactions, but without their ad hoc inclusion in the theory. In this case the TC and QCD self-energies decrease logarithmically with the momenta, which allows us to build models where ETC boson masses can be pushed to very high energies, and do not lead to undesirable flavor changing interactions. Viable TC models may be built along this line including  a necessary horizontal symmetry. The different fermionic mass scales are dictated by the different strong interactions. Pseudo-Goldstone bosons acquire large masses in this class of models.
\end{abstract}

\maketitle

\vskip 0.5cm                           

Forty years ago Weinberg~\cite{wei} and Susskind~\cite{sus} verified that the Standard Model (SM) gauge symmetry breaking could be promoted by a
new strong interaction dubbed as Technicolor (TC). In this type of symmetry breaking the role of a fundamental scalar boson, the Higgs boson, is 
substituted by a composite scalar formed by new techniquarks of the TC theory. This dynamical symmetry breaking mechanism has been reviewed in
Refs.~\cite{far,hs}. Unfortunately, the TC models and its many variations are plagued by phenomenological problems as described in Ref.~\cite{hs}.

\begin{figure}[h]
\centering
\vspace*{-0.5cm}
\includegraphics[scale=0.4]{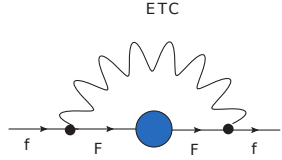} 
\vspace{-0.25cm} 
\caption[dummy0]{Ordinary fermion mass $f$ in ETC models}
\label{fig1}
\end{figure}

In TC models the ordinary fermion (f) obtains mass  coupling to technifermions (F) via extended technicolor (ETC) gauge bosons as in Fig.(\ref{fig1}). The blob in that figure is the TC self-energy, which in non-Abelian gauge theories is given by Eq.(1), where $\mu_{\tt{TC}}$ is the TC dynamical mass and $\gamma_m$  an anomalous mass dimension. 
                                             
\be
\Sigma (p^2) \propto \frac{\mu_{\tt{TC}}^3}{p^2} \(\frac{p}{\mu_{\tt{TC}}}\)^{\gamma_m}.
\label{eq1}
\ee

The generated fermion mass  is given by $m_f \propto \frac{\mu_{\tt{TC}}^3}{M_{E}^2}$. Different fermion masses appear as a consequence of different ETC gauge boson masses $M_{E}$. 
This dependence on $M_{E}$ is at the origin of all TC phenomenological problems~\cite{hs}, including the models with large $\gamma_m$ values~\cite{yama}.

\begin{figure}[h]
\centering
\includegraphics[scale=0.35]{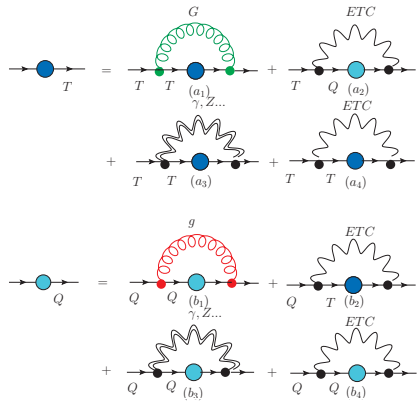} 
\vspace{-0.25cm} 
\caption[dummy0]{The coupled system of  SDEs for TC  ($T\equiv$technifermion) and QCD  ($Q\equiv$quark)  including  ETC and electroweak or other corrections. $G \,(g)$ indicates a technigluon (gluon).}
\label{fig2}
\end{figure}

\begin{figure}[h]
\centering
\includegraphics[width=0.35\columnwidth]{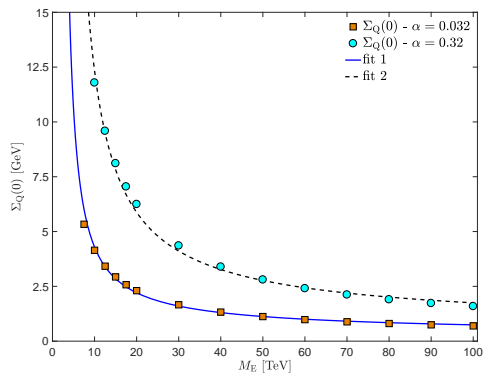}
\caption[dummy0]{$\Sigma_{\Q} (0)$ (quark running mass at $p^2=0$) as a function of $M_{\E}$ 
with the fit given by Eq.~(\ref{eqx5}). The  fit $1$ was obtained with the ETC coupling constant
{$\alpha_{\E}= 0.032$} and the fit $2$  with {$\alpha_{\E}= 0.32$}.}
\label{fig3}
\end{figure}

We have verified that when TC and QCD are coupled into a larger gauge theory (see Fig.(\ref{fig2})) the self-energies behavior are modified~\cite{us1}. The numerical solution of the self-energies where TC, described by a $SU(2)$ gauge group, is coupled to QCD is shown in Fig.(\ref{fig3}), lead to fermion masses that can be fitted by Eq.(2).

\be
m_{\Q}^{fit} = a_i [\ln (M^2_{\E}/\mu_{\tt{TC}}^2)]^{-b_i} \, ,
\label{eqx5}
\ee

In Eq.(2) the set of optimal values when the ETC coupling constant \mbox{$\alpha_{\E}=0.032$} is  
\mbox{$a_1= 203.92$ GeV} and \mbox{$b_1= 2.53$}, 
to be denoted by fit $1$. For the fit $2$, defined  when \mbox{$\alpha_{\E} = 0.32$},  we have \mbox{$a_2= 912.9$ GeV} and \mbox{$b_2= 2.82$}. The reduced chi-square of both fits are \mbox{$R^2=0.99$}. The behavior of Eq.(2) is similar to the one found in theories where bare current masses are introduced. This is not unexpected since in Ref.~\cite{us2}
we have shown analytically that at the same time that TC give masses to ordinary fermions QCD also gives masses to 
the technifermions, as well as other interactions may contribute to these masses when all their Schwinger-Dyson equations are coupled.

A most accurate description of the TC self-energy is given 
\be
\Sigma_{\T}(p^2)\approx \mu_{\tt{TC}} \left[ 1+ \delta_1 \ln\left[(p^2+\mu^2_{\tt{TC}})/\mu^2_{\tt{TC}}\right] \right]^{-\delta_2} \, .
\label{eq3}
\ee
$\delta_1$ and $\delta_2$ are parameters that depend on the QCD, TC and ETC theory. In the case with more interactions 
(e.g. electroweak) these parameters will contain corrections proportional to the charges of these
theories, and ultimately $\delta_2$ would be related to the mass anomalous
dimension of a technifermion, although its value is connected to the QCD dynamics as well as to the other interactions
present in the model.

Ordinary quark masses generated by Eq.(\ref{eq3}) will be given by~\cite{us1,us3}
\be
m_{\Q} \propto \lambda_E \mu_{\tt{TC}} [1+\kappa_1 \ln(M^2_{\E}/\mu_{\tt{TC}}^2)]^{-\kappa_2} \, ,
\label{eq4}
\ee
where $\lambda_E$ involves a ETC coupling multiplied by a Casimir operator eigenvalue and the $\kappa_i$ are also functions of the $\delta_i$  in Eq.(\ref{eq3}) as well as other possible corrections (electroweak or other interactions). The QCD self-energy has the same behavior of Eq.(\ref{eq3})
just changing $\mu_{\tt{TC}}$ by $\mu_{\tt{QCD}}$ (the QCD dynamical mass) and respective $\delta_i$ coefficients.  

Two facts are noticeable: 1) The diagrams (a2) and (b2) act as ordinary masses for technifermions and ordinary fermions (as verified in Ref.~\cite{us1} and \cite{us2}); 2) The generated masses are weakly dependent on $M_E$, and, if we neglect the logarithmic factor in Eq.(\ref{eq4}), the generated masses can be roughly described by  $\lambda_E \mu_{\tt{TC}}$ or $\lambda_E \mu_{\tt{QCD}}$ (depending on specific characteristics of the model). The small dependence on the ETC gauge boson masses allow us to promote the ETC gauge symmetry breaking at very large energies, alleviating all FCNC problems usually associated to the ETC interaction~\cite{hs}.

In the models discussed in Refs.~\cite{us3,us4} most of the fermion masses turn out to be quite heavy and proportional to $\lambda_E \mu_{\tt{TC}}$. That is the reason for introducing a horizontal (or family) symmetry, in such a way that the first fermionic generation couples only to QCD and the third generation couples preferentially to TC, generating masses of ${\cal{O}}(100)$GeV and of few MeV respectively to the third and first fermionic generation. Therefore, this class of models can explain the mass splitting between the different generations, whereas the second generation obtain mass due to      the     exchange of horizontal gauge bosons (as shown in Refs.~\cite{us3} and ~\cite{us4}).

In Refs.~\cite{us3} and \cite{us4} we have shown how a Fritzch type mass matrix can be generated 
in the TC coupled model scenario, i.e.
\br
 m_f =\left(\begin{array}{ccc} 0 & A & 0\\ A^* & 0 & B \\
0 & B^* & C
\end{array}\right),
\label{ze13} \er
where $A\approx \alpha \mu_{QCD}$ and $C\approx \alpha^\prime \mu_{\tt{TC}}$, providing a natural explanation of the 
different mass scales, where $\alpha$ and $\alpha^\prime$ are coupling constants and the $B$ entry appears due the exchange of horizontal/family gauge bosons
(see, for instance, Ref.~\cite{us4}). Eq.(\ref{ze13}) has many of the expected qualities of the known fermionic spectrum.  It is important to stress that in this scheme the TC sector will interact at leading order only with the fermions of the third generation.

\begin{figure}[t]
\centering
\includegraphics[width=0.65\columnwidth]{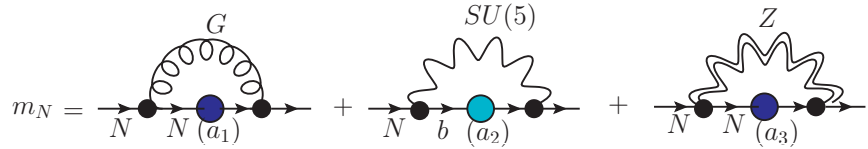}
\caption[dummy0]{ Contributions to the neutral technifermion mass in the coupled TC scheme.}
\label{fig4}
\end{figure}

Usually we can find in TC models a neutral technifermion $(N)$ with ordinary neutrino quantum numbers. This technifermion may be associated to the lightest pseudo-Goldstone boson (${\bar{N}}\gamma_5\tau^i N$, where $i$ indicate electroweak indexes). In the coupled 
model of Ref.~\cite{us4},
where TC and QCD are embedded into a $SU(5)$ gauge group, $N$ receives mass from diagrams
like the one of Fig.(\ref{fig4}), where the first diagram generates a dynamical TC mass, the second generates a ``current" mass (leading to the logarithmic
self-energy behavior), and the third one generates a weak mass correction of the following order
\be
m_N \approx g^2_w \mu_{\tt{TC}} ,
\label{eq5}
\ee
where $g_w$ is the weak coupling constant. This mass is of ${\cal{O}}(100)$GeV, which can be even enhanced by the neglected logarithmic term of Eq.(\ref{eq4}). Simple use of the 
Gell-Mann-Oakes-Renner relation will lead to this neutral pseudo-Goldstone boson a mass $M_{\Pi} \approx 150 $GeV~\cite{us4}. This pseudo-scalar boson will decay into  weak gauge bosons and may have escaped detection up to now.

These models may contain several pseudo-Goldstone bosons, but the charged and colored ones certainly will be heavier than the neutral one discussed above. It is also worth
mentioning that the scalar boson playing the role of the Higgs boson will be light in
this scenario. The reason was discussed at length in Refs.~\cite{us3,pdn}.

We conclude pointing out the relevant qualities of the TC coupled models: a) The coupled TC and QCD self-energies behavior are modified and lead to generated fermion
masses weakly dependent on the ETC mass scale, b) The mass splittings between different generations are not related to different ETC boson masses, but they appear as a consequence of a
horizontal or family symmetry. This new symmetry is necessary in order to build realistic models, c) As a consequence of the self-energy behavior described in Eq.(\ref{eq3})
pseudo-Goldstone boson masses turn out to be heavier and scalar masses lighter. It is clear that the mass of the first fermionic generation will be generated by QCD, while
the mass of the third generation will be generated by TC, explaining naturally the mass difference observed in the fermionic spectrum.

\section*{Acknowledgments}

This research  was  partially supported by Funda\c c\~ao de Amparo a Pesquisa do Estado de S\~ao Paulo (FAPESP) (A.A.N.) under the grant 2019/07821-6 and by the Conselho Nacional de Desenvolvimento Cient\'{\i}fico e Tecnol\'ogico (CNPq)
under the grants 302663/2016-9 (A.D.) and 302884/2014 (A.A.N.).

\end{document}